# In-situ tunable, room-temperature polariton condensation in individual states of a 1D topological lattice

*Ioannis Georgakilas[1,2], Rafał Mirek[1], Darius Urbonas[1], Michael Forster[3], Ullrich Scherf[3], Rainer F. Mahrt[1] and Thilo Stöferle[1]\**

[1] IBM Research Europe – Zurich, Säumerstrasse 4, 8803 Rüschlikon, Switzerland

[2] Institute of Quantum Electronics, ETH Zurich, Auguste-Piccard-Hof 1, 8093 Zürich, Switzerland

[3] Macromolecular Chemistry Group and Wuppertal Center for Smart Materials & Systems (CM@S), Bergische Universität Wuppertal, Gaußstraße 20, 42119 Wuppertal, Germany

*E-mail: tof@zurich.ibm.com

## Abstract

In recent years, exciton-polariton microcavity arrays have emerged as a promising semiconductor-based platform for analogue simulations of model Hamiltonians and topological effects. To realize experimentally a variety of Hamiltonians and change their parameters, it is essential to have highly tunable and easily engineerable structures. Here, we demonstrate in-situ tunable, room-temperature polariton condensation in individual states of a one-dimensional topological lattice, by utilizing an open-cavity configuration with an organic polymer layer. Angle-resolved photoluminescence measurements reveal the band structure of the Su-Schrieffer-Heeger chain, comprised of S-like and P-like bands, along with the appearance of discrete topological edge states with distinct symmetries. Changing the cavity length in combination with vibron-mediated relaxation in the polymer allows us to achieve selective polariton condensation into different states of the band structure, unveiled by nonlinear emission, linewidth narrowing, energy blue-shift and extended macroscopic coherence. Furthermore, we engineer the bandgap and the edge state localization by adjusting the interaction between adjacent lattice sites. Comparison to first-principles calculations showcases the precision of the polariton simulator. These results demonstrate the versatility and accuracy of the platform for the investigation of quantum fluids in complex potential landscapes and topological effects at room temperature.

# 1. Introduction

Cavity exciton-polaritons (polaritons in what follows) are hybrid, light-matter, quasiparticles resulting from strong coupling between an optical microcavity mode and an excitonic transition in a semiconductor.[1] Polaritons behave as composite bosons which, when reaching a critical density, can undergo non-equilibrium Bose-Einstein Condensation,[2] marked by the appearance of macroscopic coherence and superfluidity. Polariton condensates have been demonstrated in a variety of material systems, originally in II-VI[3] and III-V[4] compound semiconductor heterostructures, and more recently, in materials like organic[5,6] and perovskite[7] semiconductors, providing simplified processing and enabling room temperature operation. Polariton condensates have been a work horse to study fundamental phenomena such as quantized vortex formation,[8] superfluidity,[9] Kardar-Parisi-Zhang physics,[10] but also allowed realizing polaritonic devices, including polariton lasers[11] and all-optical logic.[12–14]

Arrays of coupled polariton condensates are an increasingly popular platform for analogue simulations of Hamiltonians, drawing inspiration from the field of ultracold atomic gases. By using established nanofabrication techniques, polariton lattices can be created where the hybrid light-matter nature of polariton condensates allows for shaping and controlling the photonic part of their wavefunction, while interparticle interactions are supported through their matter part. Moreover, spectroscopic techniques combined with reciprocal- and real space imaging enable direct measurement of the lattice's band structure and the amplitude of the wavefunction.[15] Motivated by this, a variety of one- (1D) [16–19] and two-dimensional (2D) [20–23] polariton lattices have been realized, allowing to study a plethora of physical effects such as topologically protected lasing,[24,25] flat band formation,[26–28] polaritonic graphene[29] and spin frustration.[30]

In this work, we investigate a polariton Su-Schrieffer-Heeger (SSH) chain, i.e. a 1D lattice comprised of adjacent sites with two alternating coupling strengths.[31] This model was originally established to mimic the alternating single and double bonds in polyacetylene[32] and, more recently, has been implemented as a basic model supporting topologically protected states. **Figure 1a** displays the two distinct configurations of a finite-length SSH chain;[33] (i) the so-called topologically "trivial" configuration (the chain starts and ends with the strong bond), which does not support topological states, and (ii) the topologically "non-trivial" configuration (the chain starts and ends with the weak bond), which induces the formation of topological edge states inside a forbidden energy gap (SSH gap in what follows) arising from the chain's staggered nature.

Many previous studies were based on monolithic cavities, where the system was engineered such that polaritons condense in a specific, desired state.[34–38] In the present work, we exploit an in-situ tunable cavity configuration, in combination with vibron-mediated polariton scattering[13,39] allowing to achieve selective polariton condensation[23] in different topologically trivial and non-trivial states. Therefore, within the very same SSH chain, we can observe the formation of two distinct topological edge state condensates, originating from two bands with different symmetries. Furthermore, we demonstrate precise control over the size and energy of the SSH gap and the localization of the topological edge states and compare it to *ab-initio* numerical calculations. Our results showcase the potential of this highly tunable and easy-to-engineer platform for conducting room temperature analogue simulations[40] and investigating topological effects.[41,42]

## 2. Results

We use a tunable open-cavity setup, which is composed of two separate halves as illustrated in Figure 1b. The fabrication and setup are detailed in the Methods section. In short, the bottom cavity half is comprised of a glass substrate, coated with a Distributed Bragg Reflector (DBR) and, on top of it, a thin layer of methyl substituted ladder-type polymer (MeLPPP) with additional encapsulation. The top cavity half consists of a DBR deposited on a glass substrate in which Gaussian-shaped deformations have been patterned, serving as potential wells for the polaritons where alternating center-to-center spacings effectively form an SSH chain. Each Gaussian deformation constitutes a single lattice site, which induces lateral confinement to the polariton wavefunction, resulting in the formation of a set of discrete energy states.[43] Additionally, engineering the spatial overlap of neighboring sites allows to control the coupling strength between them.[44] The cavity halves are mounted on separate nano-positioning stages for *XYZ* translation, enabling to tune the cavity length and, therefore, precisely control the energy of the cavity resonance. Previous experiments conducted in similar cavity configurations, patterned with a single Gaussian deformation, have shown that this sample configuration is in the strong light-matter interaction regime with a Rabi splitting of approximately $2\Omega = 100 - 150$ meV.[43,44] As the SSH band structure makes precise, direct determination of coupling strengths with coupled-oscillators models impractical, we rely on these single-site measurements to infer a similar light-matter coupling strength for the SSH chain. The detuning of the cavity from the exciton is about - 200 meV, corresponding to an excitonic fraction of ~9% in the lower polariton branch that we are considering for the polariton condensates.

## 2.1 Polaritons in an SSH chain

We study a topologically non-trivial chain configuration, comprised of 14 lattice sites, with 0.72 µm and 1.26 µm spacings for the strong and weak bond, respectively, resulting in effective couplings $J_1 = 9.2$ meV and $J_2 = 4.6$ meV (see Supplementary Information). The structural characterization of the lattice and the measurements revealing its topological nature are reported in Figure 1c-f. An Atomic Force Microscopy (AFM) measurement of the structure (Figure 1c) shows a chain length of 14 µm and a depth of 40 nm for each Gaussian deformation. We populate the band structure of the lattice with polaritons by exciting the sample below the polariton condensation threshold with an off-resonant continuous-wave (CW) laser. Exciting locally (2-4 lattice sites) and detecting the angle-resolved emission at the center of the lattice allows us to observe the well-resolved band structure comprised of $S$- and $P$-like bands and the formation of energetic bandgaps of the order of 10 meV (Figure 1d). $S$- and $P$-SSH gaps which open due to the bond alternation are indicated with the dashed lines. By repeating the measurement at the edge of the chain, the band structure exhibits two non-dispersive energy states which appear inside the SSH gaps (Figure 1e), indicating the formation of topological edge states within the polariton lattice's band structure. The observed band structure is characteristic for a linear SSH chain and can be theoretically reproduced by solving the Schrödinger equation for the full potential in real space (Figure 1f) (see Supplementary Information).

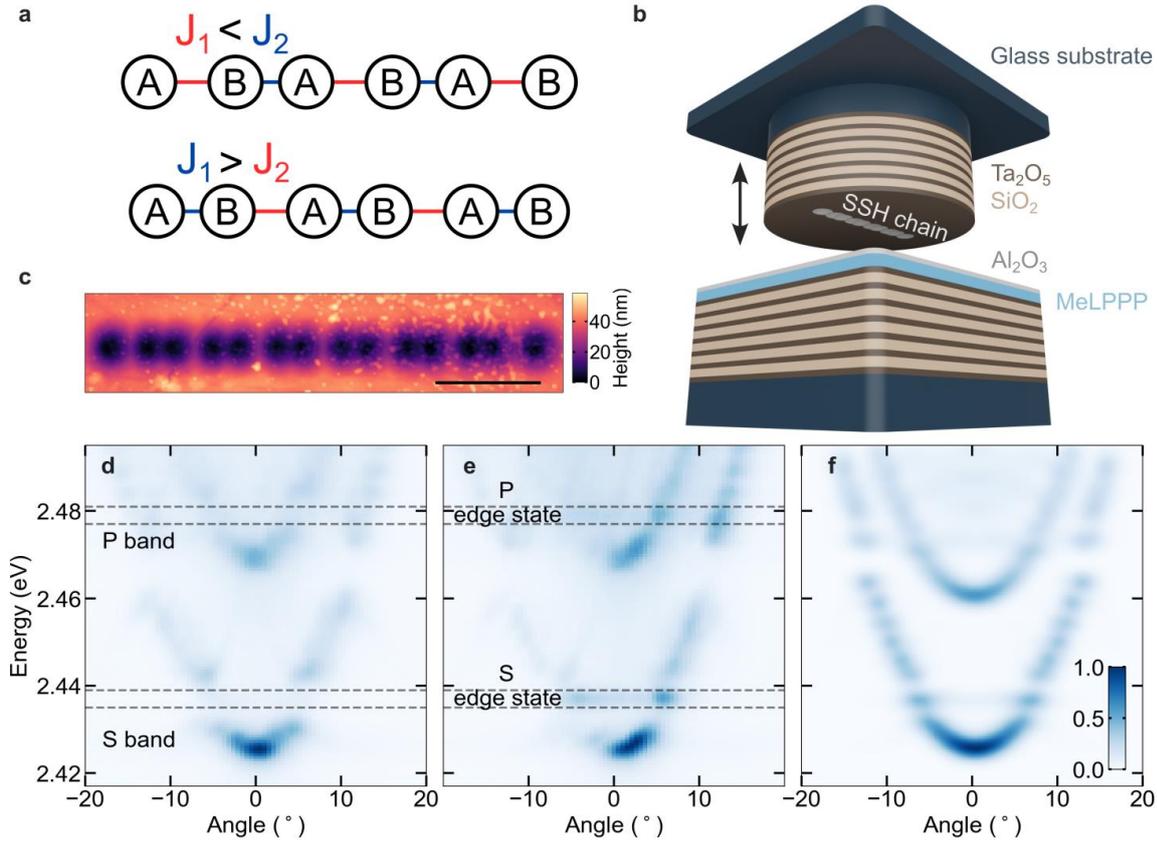

**Figure 1.** Polaritons in an SSH chain. a Sketches of the topologically non-trivial (top) and trivial (bottom) SSH configurations. Large spacing and red color indicate the small effective coupling, whereas small spacing and blue color indicate the large effective coupling. b Illustration of the employed tunable cavity comprised of two separate halves. The vertical black arrow indicates that the distance between the two halves can be tuned. The bottom half comprised the organic layer (blue color) on top of a DBR (alternating layers). The top part consists of a patterned DBR where the grey colored structure indicates the patterned SSH chain. c AFM image of the studied structure. Scale bar equals 3 µm. d,e Angle-resolved photoluminescence measurements of the SSH chain performed when exciting at the middle (d) and at the edge (e) of the structure. The band structure is comprised of S- and P-like bands (d) and two SSH gaps which appear due to the bond alternation. Measuring the structure's edge reveals the topological edge states inside the SSH gaps (e). Two dashed lines indicate the region of the SSH gaps and clearly visualize the absence of the edge states in (d) and their appearance in (e). f Simulated band structure of the SSH chain.

## 2.2 Selective polariton condensation in different lattice modes

Next, we studied the emission of the structure above polariton condensation threshold, as displayed in **Figure 2**. Upon non-resonant excitation of the organic polymer MeLPPP, hot excitons are formed, that relax predominantly through internal conversion, forming an excitonic reservoir. A single-step relaxation scheme including a molecular vibron enables efficient population of polariton states one vibron energy below the excitonic reservoir, avoiding inefficient multi-step relaxation processes.[45,46] Therefore, in our organic polymer, polariton condensation occurs efficiently when the detuning between the polaritonic mode and the exciton reservoir matches a strong vibronic transition (in our material 200 meV) (Figure 2a). When exciting locally (2 – 4 sites) at the lattice's edge, polariton condensation in the *S*-band topological edge state has been observed, displaying the expected condensate signatures of nonlinear increase of the emission intensity, full-width at half-maximum (FWHM) narrowing and energy blue-shift, with a condensation threshold between 100 – 200 μJ cm$^{-2}$ (Figure 2b). The value of the observed threshold is consistent with the reported polariton condensation threshold of 130 μJ cm$^{-2}$ in the ground state of a Gaussian-shaped cavity with the same active layer.[43] By tuning the cavity length to energetically shift the band structure and utilizing the vibron-mediated condensation process, we can trigger polariton condensation in different topologically non-trivial and topologically trivial states of our structure (Figure 2c-h). The two experimental parameters needed to adjust for the mode-selective condensation are the cavity length and the position of excitation. Exciting at the edge of the lattice, allows to selectively condense into either the *S*- or the *P*-edge states (Figure 2d,g), whereas when exciting at the center of the chain, one can change between the binding and anti-binding *S*- and *P*-band (trivial) bulk modes (Figure 2c,e,f,h) by adjusting the resonance.

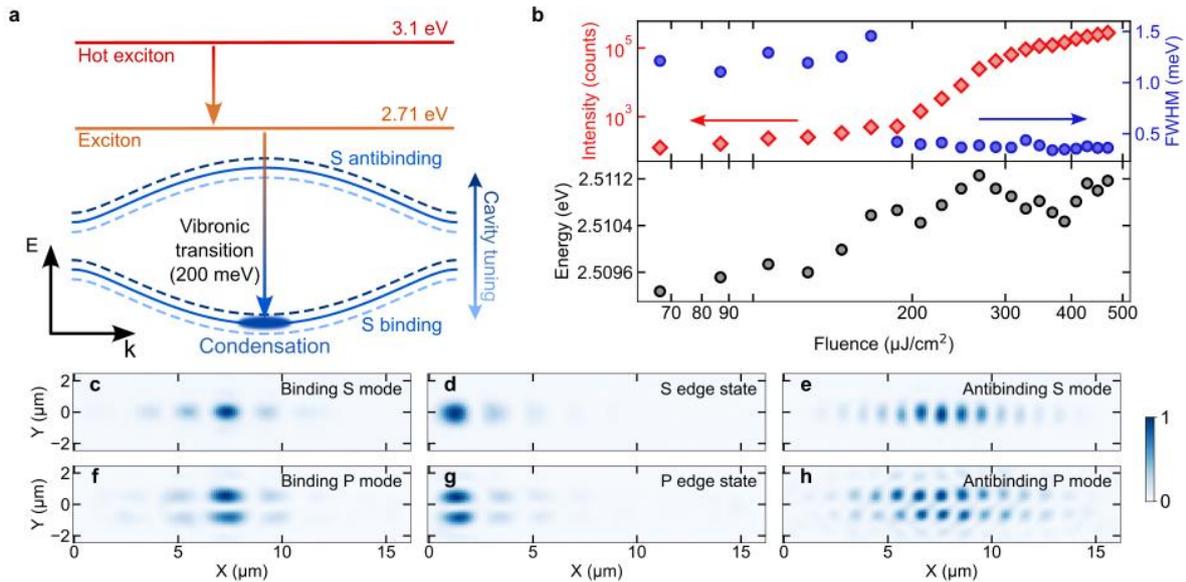

**Figure 2.** Tunable polariton condensation. a Sketch of the vibron-mediated polariton condensation process. We create hot excitons (red) at a photon energy of 3.1 eV by driving the system non-resonantly. Through internal conversion, the hot excitons relax and form the exciton reservoir (orange) at 2.71 eV. The displayed dispersion (blue) is typical for the folded S-band of the SSH chain. Tuning the cavity length shifts the dispersion in energy (dashed lines), bringing different parts of it in resonance with the 200 meV vibronic transition and therefore inducing efficient polariton condensation in different lattice modes. b Excitation fluence dependent measurements of the intensity (top panel, red diamonds, left axes), FWHM (top panel, blue dots, right axes) and blue-shift (bottom panel). c-h Real space images of different topologically non-trivial (d, g) and topologically trivial (c, e, f, h) lattice modes in which one can selectively condense. Each real space image is normalized, and the emission intensity is mapped by the colour bar at the very right.

As extended macroscopic coherence is a hall mark of polariton condensation, we used a Michelson interferometer setup with a retroreflector mounted in one arm to study the first-order coherence of the polariton condensates. The results are presented for the edge state mode and the binding and anti-binding modes of the *P*- band manifold in **Figure 3**. The interferograms show that the condensates exhibit spatial coherence extending over nearly the whole structure (Figure 3a). The highest intensity of the condensate appears at the edges in the case of the edge state (Figure 3a left panel) and at the center for the bulk modes (Figure 3a middle and right panels), due to local excitation at the edge and the middle of the structure, respectively. Subsequently, tuning the length of one of the two interferometer arms, which changes the effective time delay $\Delta t$, allows to measure the temporal behavior of the condensate coherence by extracting the fringe visibility of the interferogram, resulting in a Gaussian autocorrelation function with a FWHM of 4.3 ps (Figure 3b). Such a Gaussian envelope for the phase coherence can be attributed to number fluctions and polariton interactions.[47] The observed coherence time is an order of magnitude longer than the polariton lifetime in the system and is comparable to the excitation pulse duration, which effectively determines the lifetime of the polariton condensate.

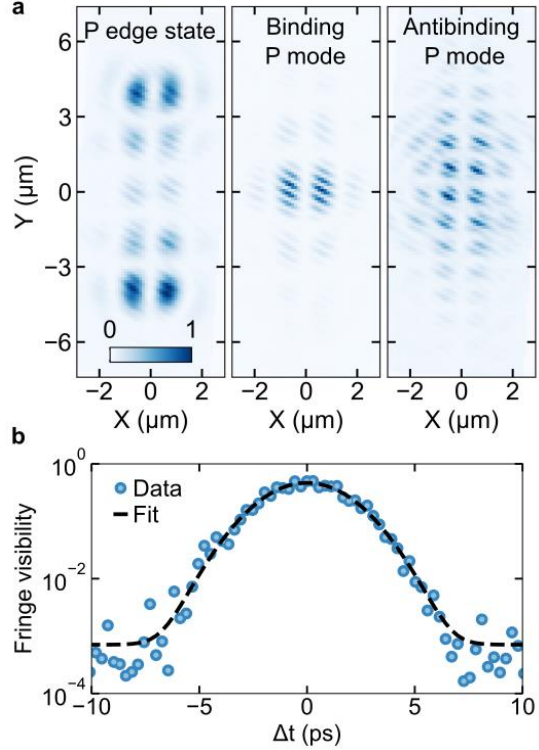

**Figure 3.** Condensate spatial and temporal first order coherence. a Interferometric measurements of the three P lattice mode condensates. For each measurement, the interference fringes extend through the whole condensate. Each image has normalized intensity which is described by the color bar inset at the bottom of the edge state image. b Measurement of the temporal coherence of the binding P mode condensate. The blue dots are the experimental data extracted by Fourier transformation of the real space interferograms. The dashed black line represents a Gaussian fit to the data with a FWHM of 4.3 ps.

### 2.3 SSH gap and edge state engineering

In view of an analogue polariton simulator, the developed structure should be easily engineerable to realize varying parameters of the Hamiltonian under study. By measuring three separate chains with different coupling strengths, we assess the impact of the interaction between sites on the topological features and the band structure of the lattice, as shown in **Figure 4**. The three investigated chains share a common strong bond coupling ($J_1 = 9.2$ meV), while they have different weak bonds ($J_2 = 4.6$ meV, 3.2 meV and 2.0 meV). The angle-resolved measurements show that for larger site spacing (weaker coupling), corresponding to an enhanced coupling contrast between weak and strong bond, the SSH gap becomes larger, in line with theory predictions[33] (Figure 4a). Moreover, by studying the real space emission of the condensed edge state in the three chains with different bandgap sizes, we can demonstrate the effect of the bandgap on the localization of the edge state (Figure 4b). As expected from theory,[33] the larger the SSH bandgap is, the more isolated is the edge state with respect to the bulk modes, leading to a tighter spatial localization at the edge of the lattice (Figure 4b, top panels). Additionally, by fitting the mode profiles with a sinus-squared modulation with an exponentially decaying envelope, we extract an exponential decay length $\tau$ for each state (Figure 4b bottom panels).

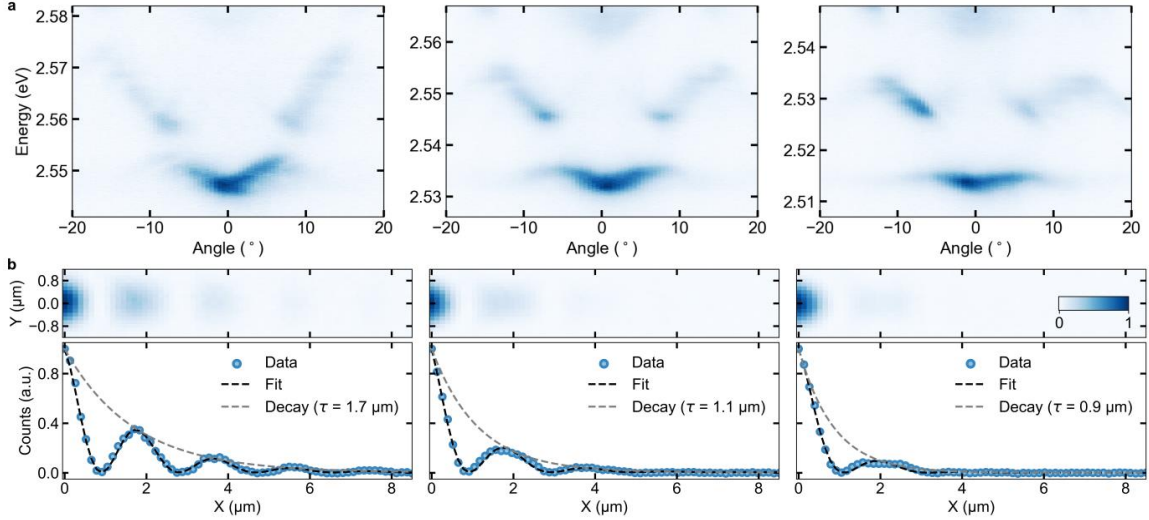

**Figure 4.** SSH gap and topological edge state engineering. a Angle-resolved photoluminescence measurements of three structures with varying weak bonds (from left to right; 4.6 meV, 3.2 meV, and 2.0 meV), showing the evolution of the bands and the bandgap of an SSH chain in topologically trivial configuration. b The top panels show the real space mode profiles of the condensed S-band edge states for three non-trivial structures with matching bandgaps as in (a), while the bottom panels show the fitted line profiles of the three edge states with their respective calculated decay lengths (intensity counts were integrated over the vertical real space image axis). All the images (a and b) are normalized and are described by the colour bar inset in the top right part of b.

**Figure 5** summarizes the direct effect of the engineered coupling to the SSH gap variation and the edge state localization and compares the experimental results to the numerical simulations (for additional details see SI). The measured width of the SSH gap (Figure 5a, blue points) increases in a slightly nonlinear way versus the difference between the two couplings $J_1$ and $J_2$. This increasing trend is matched well by simulations, where we solve the Schrödinger equation (black line), in contrast to a perfectly linear trend predicted by a simple tight-binding model[33] (grey dashed line). The experimentally obtained decay lengths of the three different edge states (Figure 5b, blue points) are also plotted versus the difference between the two couplings. The larger SSH gap results in more confined edge states, again observing a good match between experiment and simulations.

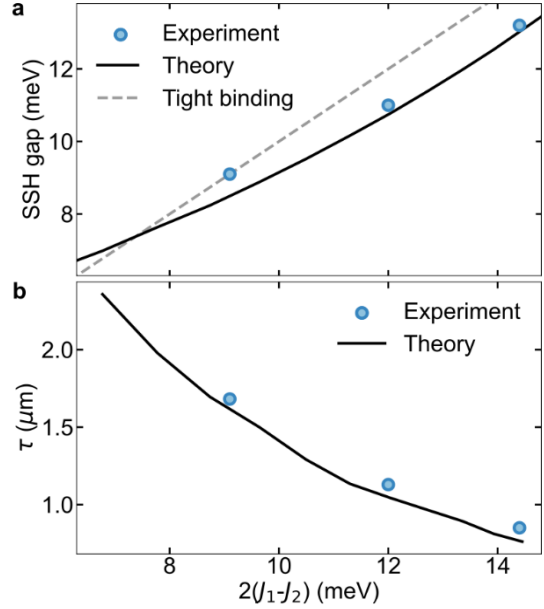

**Figure 5.** SSH gap and edge state decay versus $2(J_1-J_2)$. a Plotted experimental (blue dots) and simulated (black line) SSH gap width versus $2(J_1-J_2)$. The observed SSH gap's slightly nonlinear increase is matched well with the numerical simulations. b Extracted from the fit of the spatial images (Fig. 4), experimental (blue dots) and simulated (black line) decay lengths versus $2(J_1-J_2)$. Experiment and theory agree well, and both illustrate the tighter confinement of the topological edge state when it resides in a larger SSH gap.

## 3. Conclusion

We have presented a facile and highly tunable platform allowing to study organic, room temperature polariton condensation in topologically protected states. The topological edge states have been implemented by a 1D cavity array, mapping to an SSH chain model. Measurements below condensation threshold illustrated a band structure with *S*- and *P*-band manifolds, which, when excited locally at the edge of the chain, displayed the formation of two edge states, confirming the topologically non-trivial nature of the structure. We demonstrated polariton condensation in the *S*-edge state by driving the system above threshold while pumping spatially at the edge of the structure. Vibron-assisted relaxation together with the tunable cavity configuration allowed us to selectively condense polaritons in different lattice modes, showing polariton condensation in two distinct topological edge states originating from bands with different symmetry as well as in different bulk states. All condensates exhibited spatially and temporally extended coherence, revealed by interferometric measurements using a Michelson interferometer.

Furthermore, we engineered the SSH gap and thereby altered the edge state localization by controlling the coupling strength within the subunits of the SSH chain. The formation of a larger gap for enhanced coupling-strength contrast results in a tighter confined edge state. The fact that our experiments accurately match first-principles calculations emphasizes the quality and faithfulness of the analogue simulations. These results manifest the high degree of engineerability and tunability of our system, which holds great potential for exploring complex potential landscapes and topological properties within the realm of photonics as well as many-body physics with bosonic quantum fluids at ambient conditions.

## 4. Methods

*Fabrication*: We use a tunable, open-cavity setup, with a resonator comprised of two separate halves. Both "half cavities" are mounted on *XYZ* nano-positioning stages to tune the distance between them plus tilting degrees of freedom to allow for parallel alignment. The "top" half has been fabricated by performing optical lithography and wet etching with concentrated HF to create a ~30 µm tall and ~200 µm wide mesa structure in the center of a glass substrate (1 cm x 1 cm). The mesa reduces the effective surface area of the two approaching cavity halves, therefore strongly decreasing the sensitivity to particle contamination inside the tunable resonator, allowing the two parts to approach routinely on a hundred nm scale. On top of the mesa's surface, we used Focused Ion Beam (FIB) milling to pattern several 1D arrays of spatially overlapped Gaussian deformations with alternating center-to-center spacings, equivalent to SSH chains. By means of Ion Beam Deposition (IBD) 6.5 quarter-wave layer pairs of $SiO_2/Ta_2O_5$ have been deposited to fabricate a DBR which retains the morphology of the underlying substrate/pattern. For the "bottom" half, a 35 nm-thick methyl substituted ladder type polymer (MeLPPP[48]; $Mn$ = 31,500 (number averaged molecular weight), $Mw$ = 79,000 (weight averaged molecular weight)) film is deposited by spin-coating from an 1%-weight toluene solution on a flat glass substrate comprising another DBR mirror with 6.5 layer pairs with an additional 20 nm $SiO_2$ spacer. The polymer film is protected from photodegradation by encapsulation with a 20 nm thick $Al_2O_3$ layer, deposited by means of an electron beam evaporator.

*Optical Characterization:* Measurements below condensation threshold have been conducted with a continuous-wave, 405 nm laser diode, coupled to a single-mode fiber. To drive the system to the condensation regime, a frequency-doubled, amplified Ti:sapphire laser at 400 nm with a 1 kHz repetition rate and approximately 150 fs pulse duration was coupled into a single-mode photonic crystal fiber. The excitation was focused on the sample by a 100x microscope objective with numerical aperture (NA) = 0.5, resulting in beam sizes of around 1.5 – 3 μm. The same objective is used for collecting the signal corresponding to the different mode profiles and condensate threshold measurement of Figure 2. For the angle-resolved and interferometric measurements, we collect the light exiting from the bottom cavity half with a 20x objective with NA = 0.5. For the angle-resolved dispersion imaging, the signal is sent to the front entrance of a 0.5 m-long monochromator (with 300 and 1800 lines/mm gratings) coupled to a liquid nitrogen-cooled camera detector, with the lattice structures aligned parallel to the slit. For the first-order coherence measurements in Figure 3, the signal is instead sent to a Michelson interferometer with a corner retroreflector in the adjustable arm path.

**Acknowledgements**

We thank the team of the IBM Binnig and Rohrer Nanotechnology Center, Daniele Caimi and the Quantum Photonics team for support with the sample fabrication. We acknowledge funding from EU H2020 EIC Pathfinder Open project "PoLLoC" (Grant Agreement No. 899141), EU H2020 EIC Pathfinder Open project "TOPOLIGHT" (Grant Agreement No. 964770) and EU H2020 MSCA-ITN project "AppQInfo" (Grant Agreement No. 956071).

**ToC figure:**

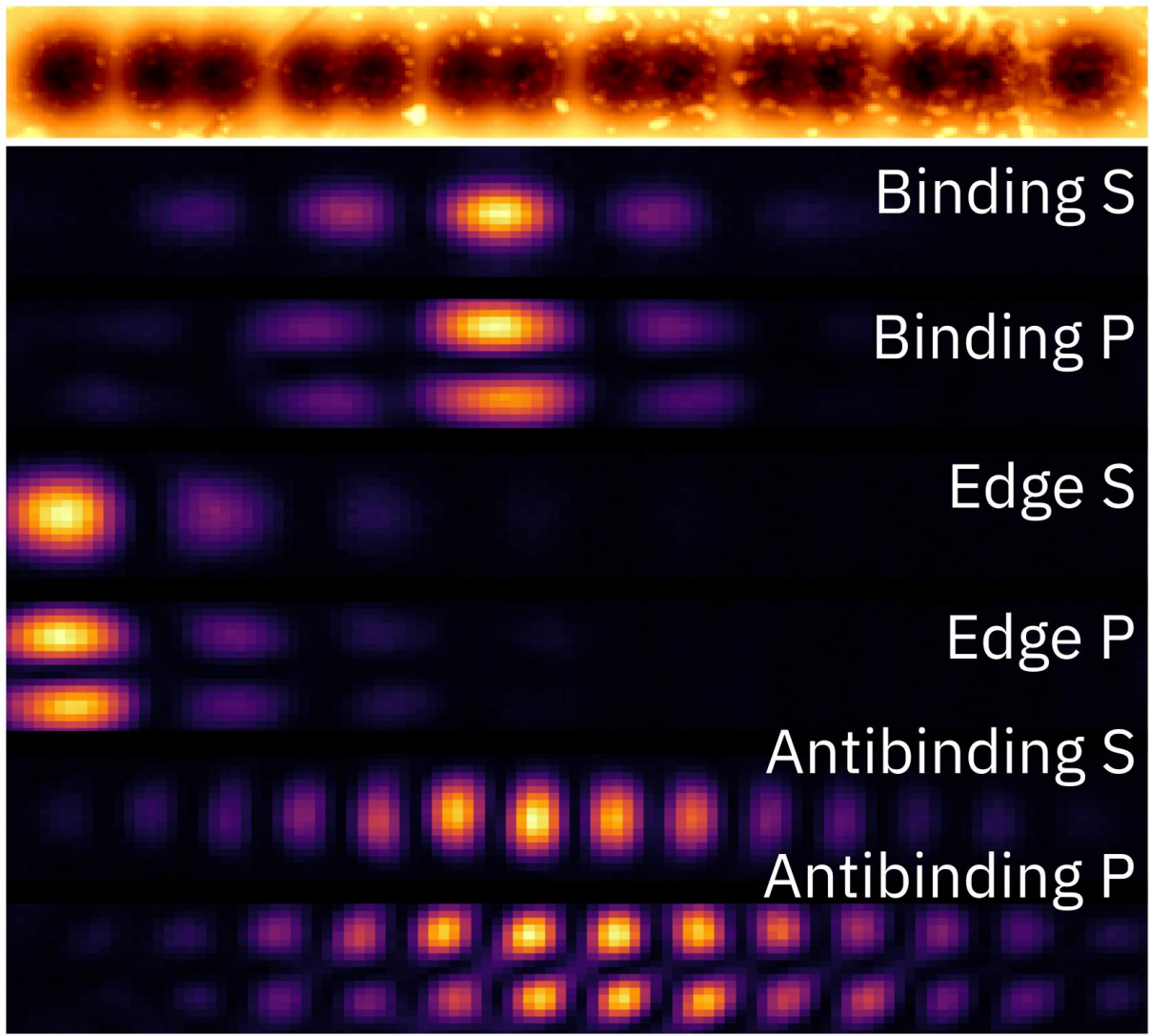

Binding S

Binding P

Edge S

Edge P

Antibinding S

Antibinding P

**ToC text:**

Topological and non-topological states of a Su-Schrieffer-Heeger chain are studied using exciton-polaritons in a nanostructured, length-tunable microcavity. Selective polariton condensation into distinct edge modes and different lattice modes is demonstrated and compared to first-principles calculations. These results showcase the versatility and the veracity of this polariton simulator platform for investigating quantum fluids in complex potential landscapes.

# Supporting Information

**In-situ tunable, room-temperature polariton condensation in individual states of a 1D topological lattice**

*Ioannis Georgakilas[1,2], Rafał Mirek[1], Darius Urbonas[1], Michael Forster[3], Ullrich Scherf[3], Rainer F. Mahrt[1] and Thilo Stöferle[1]\**

[1] IBM Research Europe – Zurich, Säumerstrasse 4, 8803 Rüschlikon, Switzerland

[2] Institute of Quantum Electronics, ETH Zurich, Auguste-Piccard-Hof 1, 8093 Zürich, Switzerland

[3] Macromolecular Chemistry Group and Wuppertal Center for Smart Materials & Systems (CM@S), Bergische Universität Wuppertal, Gaußstraße 20, 42119 Wuppertal, Germany

\*E-mail: tof@zurich.ibm.com

**Theoretical model – two-dimensional Schrödinger equation**

To describe the observed effects, we perform a first-principles calculation using an effective Hamiltonian for 2D massive particles:

$$\hat{H} = \frac{\hbar^2 \mathbf{k}^2}{2 m_{eff}} + V(\mathbf{r})$$

where $\mathbf{k}$ corresponds to the wavevector in 2D momentum space, $m_{eff}$ is the effective mass and $V(\mathbf{r})$ is the 2D potential formed by the SSH chain of Gaussian-shaped defects. We diagonalize the Hamiltonian to obtain the eigenmodes and eigenvalues. The dispersion relations were obtained by calculating the Fourier transform of the eigenmodes and assigning them a Gaussian width, matching the experimentally observed value of around 4 meV.

**Extraction of the simulation parameters and estimation of effective coupling $J$**

First, we need to extract the value of the effective mass ($m_{eff}$) and the potential depth ($V_{min}$), that corresponds to a physical depth of 40 nm, to use in the previously described model.

Therefore, we measured the spectra of four pairs of coupled Gaussian deformations (Figure S1a) that have the same site-to-site distance as the four bonds used for different SSH chains in the main experiment and consequently result in same effective coupling $J$. We fitted the model to the doublet measurements, and we extracted $m_{\mathrm{eff}} = 2 * 10^{-5} m_{\mathrm{e}}$ with $m_{\mathrm{e}}$ being the electron mass and $V_{\mathrm{min}} = 80$ meV. The simulated spectra, results of the fitting, are displayed in Figure S1b.

To determine the effective couplings $J$, we used again the four experimental spectra of the coupled Gaussian deformations. Two coupled Gaussian deformations act as a polaritonic molecule, where evanescent coupling of the polariton wavefunctions leads to mode splitting. The resulting splitting is equal to twice the effective coupling $J$ of the polaritonic molecule. For the weakest and strongest bonds used for the experiments, we measured a splitting equal to 3.9 meV and 18.3 meV, respectively (Figure S1a).

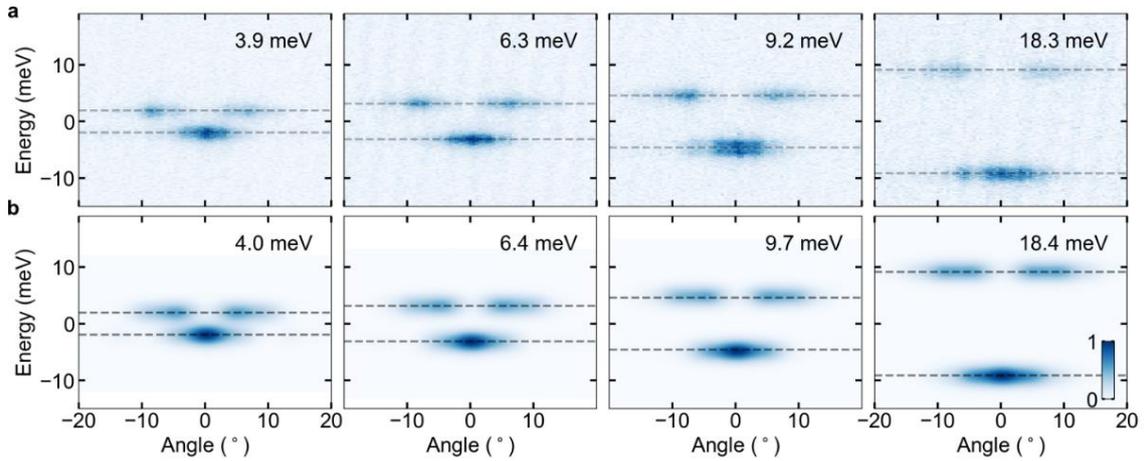

**Figure S1.** Extraction of the simulation parameters from doublet measurements. a Angle-resolved photoluminescence of coupled Gaussian deformations with center-to-center distance increasing from left to right: 0.72 μm, 1.08 μm, 1.26 μm, 1.44 μm. b Fitted angle-resolved spectra of doublets with the same spacings. The center of the y axes corresponds to the center of the gap. The dashed lines show the positions of the binding and anti-binding S-like states extracted from the experimental data. The measured and calculated gaps are illustrated in the top right corner of each panel. All the images are normalized and described by the colour bar inset in the bottom right part of last panel.

Increasing the distance between the Gaussian deformations leads to a smaller overlap of the wavefunctions and therefore to an exponentially reduced effective coupling. This can be seen from the simulated mode splittings over the large range of distances, presented in Figure S2.

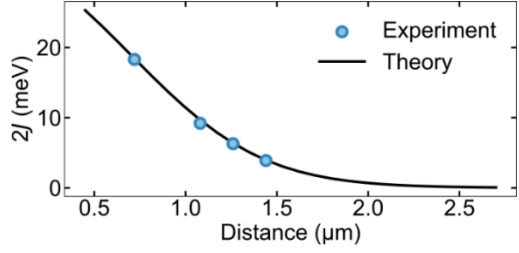

**Figure S2.** Coupling strength over distance for coupled Gaussian deformations. Measured (blue points) and calculated (black line) coupling strength over center-to-center distance between two Gaussian deformations.

**Band structure simulation and comparison to experiment**

Here, after obtaining all the required parameters from the fit to the coupled Gaussian deformations spectra, we use the model Hamiltonian to simulate the band structure of trivial and non-trivial chains corresponding to different effective couplings for the weak bond of the SSH chain, presented in Figure 4.

Both the experimental and simulated results are summarized in Figure S3. The top panels (Figure S3a) show the three experimentally measured band structures of Figure 4, while the remaining six panels are numerical simulations of the trivial (Figure S3b) and non-trivial (Figure S3c) configurations of structures with same number of sites and site-to-site spacing as the ones used for the experiment. All the simulated band structures are calculated such that the middle of the SSH gap resides at 2.51 eV, where polariton condensation is most efficient.

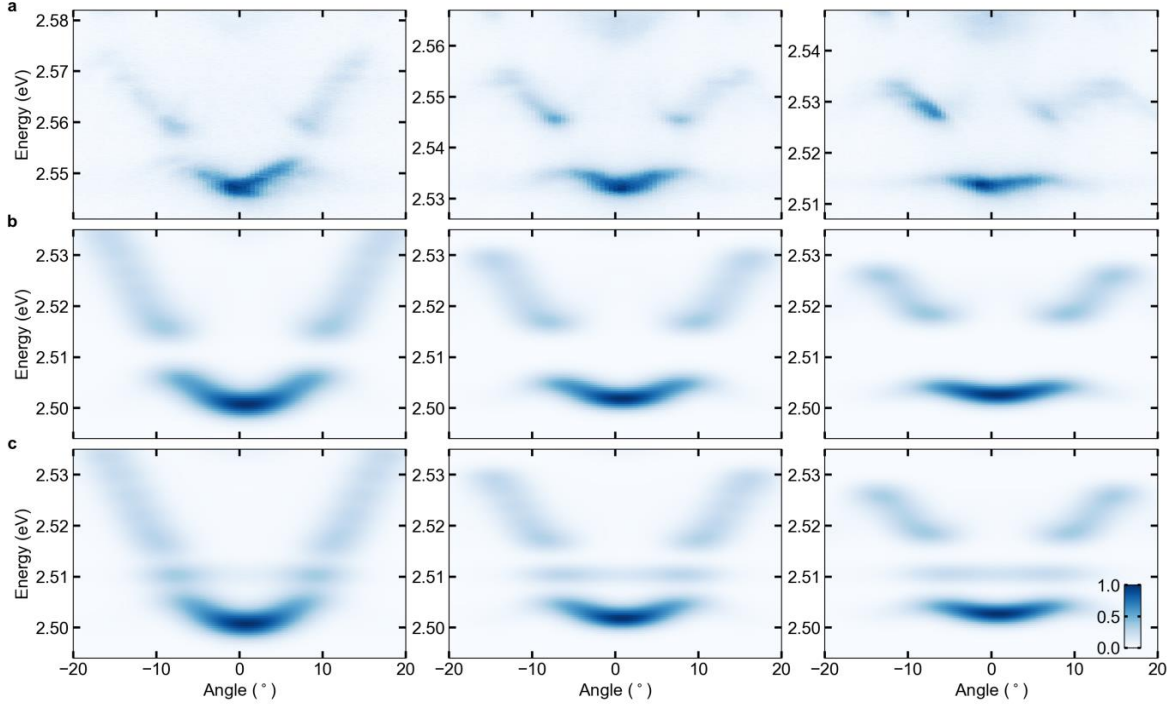

**Figure S3.** Polariton bandstructure for an SSH chain. a Angle-resolved photoluminescence measurements of three trivial SSH structures with varying weak bonds. Calculated dispersions for trivial (b) and non-trivial (c) SSH chains. The calculated dispersions have the center of the SSH gap at 2.51 eV where the polariton condensation of the S edge state occurs. Left, center and right panels correspond to center-to-center distance of the weak bond equal to 1.08 μm, 1.26 μm and 1.44 μm respectively. All the images are normalized and described by the colour bar inset in the bottom right part of last panel.

## Simulation of the edge state localization and comparison to the experiment

The comparison between the experimental and simulated edge state condensates for three different structures with different bandgaps is displayed in Figure 4. Figure S4a shows the experimentally measured edge state condensates for three chains with different bandgaps (presented in Figure 4 of main text), while Figure S4b shows the respective simulations. The localization length $\tau$ for the simulated data was extracted by fitting an exponential decay function to the local maxima.

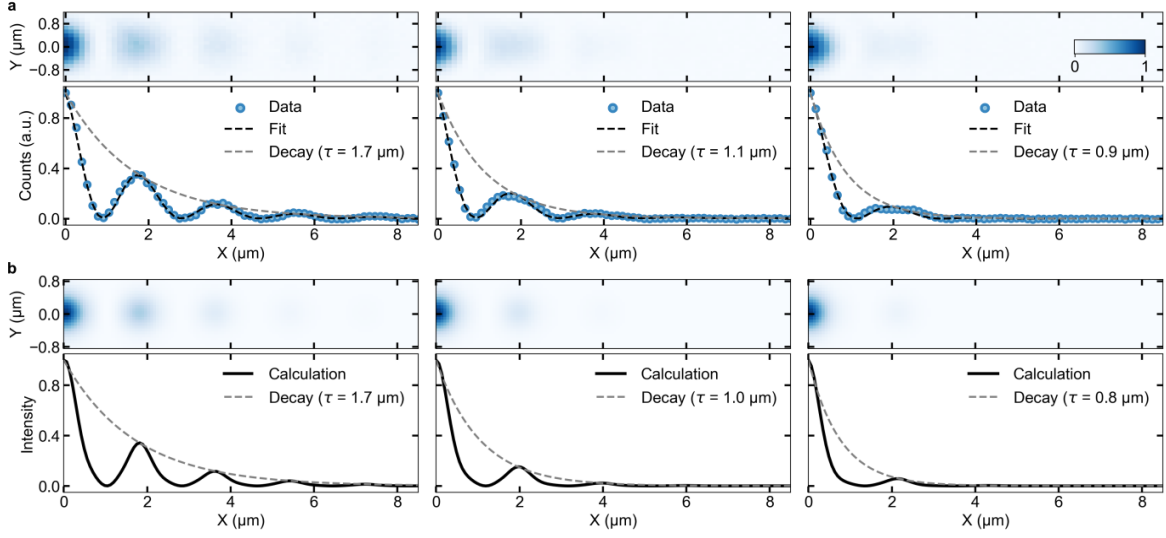

**Figure S4.** Topological edge state engineering. Measured (a) and calculated (b) real space mode profiles (top panels) of the condensed S-band edge states for three non-trivial SSH chains and the fitted line profiles (bottom panels) with their respective calculated decay lengths. Left, center and right panels correspond to center-to-center distance of the weak bond equal to 1.08 µm, 1.26 µm and 1.44 µm respectively. All the images are normalized and described by the colour bar inset in the top right panel.